\documentclass[12pt]{iopart}
\begin{document}

\title[Brown, Lee, Rho, \& Shuryak]{
The Instanton Molecule Liquid and\\ ``Sticky Molasses" Above $T_C$}

\author{G. E. Brown\dag, C.-H. Lee\ddag, M. Rho\S\ and E. Shuryak\dag  
}

\address{\dag\ Department of Physics and Astronomy,
     State University of New York, Stony Brook, NY 11794, USA
}
\eads{\mailto{Ellen.Popenoe@sunysb.edu,shuryak@tonic.physics.sunysb.edu}}

\address{\ddag\ Department of Physics,
and Nuclear physics \& Radiation technology Institute (NuRI),
Pusan National University, Busan 609-735, Korea
}
\eads{\mailto{clee@pusan.ac.kr}}

\address{\S\ Service de Physique Th\'eorique, CEA/DSM/SPhT. Unit\'e de
recherche associ\'ee au CNRS, CEA/Saclay, 91191 Gif-sur-Yvette
c\'edex, France,\\
and Department of Physics, Hanyang University, Seoul 133-791, Korea
}
\eads{\mailto{rho@spht.saclay.cea.fr}}

\begin{abstract}
The main objective of this work is to explore the evolution in the
structure of the quark-antiquark bound states in going down in the
chirally restored phase from the so-called ``zero binding points"
$T_{zb}$ to the QCD critical temperature $T_c$ at which the
Nambu-Goldstone and Wigner-Weyl modes meet. In doing this, we
adopt the idea recently introduced by Shuryak and Zahed for
charmed $\bar c c$, light-quark $\bar q q$ mesons  
$\pi, \sigma, \rho, A_1$ and gluons that at $T_{zb}$,  the
quark-antiquark scattering length goes through $\infty$ at which
conformal invariance is restored, thereby transforming the matter
into a near perfect fluid behaving hydrodynamically, as found at
RHIC. We show that the binding of these states is accomplished by
the combination of (i) the color Coulomb interaction,
(ii) the relativistic effects, 
and (iii) the interaction induced by the instanton-anti-instanton
molecules. The spin-spin forces turned out to be small. While near
$T_{zb}$ all mesons are large-size nonrelativistic objects bound
by Coulomb attraction, near $T_c$ they get much more tightly
bound, with many-body collective interactions becoming important
and making the $\sigma$ and $\pi$ masses approach zero (in the chiral limit).
The wave function at the origin grows strongly with binding, and
the near-local four-Fermi interactions induced by the instanton
molecules play an increasingly more important role as the
temperature moves downward toward $T_c$.
\end{abstract}


\submitto{\JPG}

\maketitle


\def\lsim{\mathrel{\rlap{\lower3pt\hbox{\hskip1pt$\sim$}}
     \raise1pt\hbox{$<$}}} 
     \def\gsim{\mathrel{\rlap{\lower3pt\hbox{\hskip1pt$\sim$}}
          \raise1pt\hbox{$>$}}} 
\def\be{\begin{eqnarray}}
\def\ee{\end{eqnarray}}

The concept that hadronic states may survive in the high
temperature phase of QCD, the quark-gluon plasma \cite{BLRS}, 
has been known for some time. In particular, it
 was explored by Brown et
 al.\cite{BBP91,BJBP93}. The properties of (degenerate) $\pi$ and $\sigma$
 resonances above $T_c$  in the context of the
 NJL model was discussed earlier by Hatsuda
 and Kunihiro\cite{Hatsuda_sigma}, 
  and in the instanton liquid model by Schafer and Shuryak \cite{SS_survive}.
   Recently,
   lattice calculations \cite{datta02,hatsuda2003} have shown
   that, contrary to the original suggestion by Matsui and Satz  \cite{MS},
   the lowest charmonium states $J/\psi,\eta_c$ remain bound well above $T_c$.
   The estimates of the zero binding temperature for
   charmonium
    $T_{J/\psi}$ is now limited to the interval
     $2T_c > T_{J/\psi} > 1.6 T_c$, where $T_c\approx 270\, MeV$
      is that for quenched QCD.
       Similar results for
        light quark mesons exist but are less quantitative at the moment.
	However since the ``quasiparticle" masses close to $T_c$ are
	large, they must be similar to  those for charmonium states.

  In the chiral limit
  all states above
   the chiral restoration go into
    chiral multiplets. For quark quasiparticles this is also true,
    but although the chirality is conserved during their propagation,
    they are not massless and move slowly near $T_c$ where their
    ``chiral mass'' $m=E(p\rightarrow 0)$ is large ($\sim 1$ GeV).

RHIC experiments have found that hot/dense matter at temperatures above the
critical value $T_c\approx 170 \, MeV$ is $not$ a weakly
interacting gas of quasiparticles, as was widely expected.
Instead, RHIC data have demonstrated the existence of  very robust
collective flow phenomena, well described by ideal hydrodynamics.
Most decisive in reaching this conclusion was the early
measurement of the elliptic flow which showed that equilibration
in the new state of matter above $T_c$ set in in a time $< 1$ fm/c
\cite{hydro2001,kolb}. Furthermore, the first viscosity estimates
\cite{Teaney2003} show  surprisingly low values, suggesting that
this matter is the most perfect liquid known. Indeed,  the ratio
of shear viscosity coefficient to the entropy is only $\eta/s\sim 0.1$, 
two orders of magnitude less than for water.
Furthermore, it is comparable to predictions in the infinite coupling limit
(for $\cal N$=4 SUSY YM theory)  $\eta/s = 1/4\pi$ \cite{PSS}, perhaps the
lowest value possible.

Shuryak and Zahed\cite{shuryak2003} 
have recently connected these two issues
together. They have
 suggested that large rescattering cross sections apparently present
  in hot matter at RHIC
   are generated by resonances near the zero-binding lines.
    Indeed,  at the point of
    zero binding  the scattering
    length $a$ of the two constituents goes to $\infty$ and this
    provides
    low viscosity. This phenomenon is analogous to the  elliptic flow
    observed in the expansion of trapped $^6$Li atoms rendered
    possible by tuning the scattering length to very large values via
    a Feshbach resonance~\cite{Li6,Bourel}.

Near the zero-binding points, to be denoted by $T_{zb}$, introduced by 
Shuryak and Zahed\cite{shuryak2003}
the binding is small and thus
the description of the system can be simple and nonrelativistic.
The binding comes about chiefly
from the attractive Coulomb color electric field, as evidenced in
lattice gauge calculation of Karsch and
collaborators\cite{datta02,petreczky02}, and Asakawa and
Hatsuda\cite{hatsuda2003}. The instanton
molecule interactions are less important
at these high temperatures ($T \sim 400$ MeV).

In this work we wish to construct the link between the chirally
broken state of hadronic matter below $T_c$ and the chirally
restored mesonic, glueball state above $T_c$.
Our objective  is to understand and to
work out in detail what exactly happens with hadronic states
at temperatures between $T_c$ and $T_{zb}$.
One important new point is that these chirally restored
hadrons are so small that the color charges are
locked into the hadrons at such short distances ($< 0.5$ fm)
that the Debye screening
is unimportant. This is strictly true at $T \gsim T_c$,
where there is very little free charge.
In this temperature range the nonrelativistic treatment of 
Shuryak and Zahed\cite{shuryak2003} should be changed to a relativistic one.

The relativistic current-current interaction, ultimately related with
the classical Ampere law,
is about as important as the Coulomb one,
effectively doubling the attraction.
We also found that the spin-spin
forces \cite{BLRS} are truly negligible.
In effect, with the help of the instanton molecule interaction,
one can get the bound quark-antiquark states down in energy, reaching
the massless $\sigma$ and $\pi$ at $T_c$,
so that a smooth transition can be made with the chiral
breaking at $T<T_c$.

The non-pertubative interaction from the
instanton molecules becomes very important.
Let us remind the reader of the history of the issue.
The nonperturbative gluon condensate, contributing to
the dilatational charge or trace of the stress tensor
$T_{\mu\mu}=\epsilon-3p$,
is not melted at $T_c$. In fact more than half of the vacuum gluon condensate
value remains at $T$ right above $T_c$.
the hard glue or epoxy which
explicitly breaks scale invariance but is
unconnected with hadronic masses. The rate at which the epoxy
is melted can be measured by lattice gauge simulations, and this
tells us the rate at which the instanton molecules are broken
up with increasing temperature.

As argued by Ilgenfritz and Shuryak \cite{IS},
this phenomenon can be
explained by breaking of the instanton ensemble into instanton
molecules with zero topological charge. Such molecules generate a
new form of effective multi-fermion effective interaction similar
to the orignal NJL model\cite{BLRS}.
Brown et al.\cite{bglr2003} obtained the
interaction induced by the instanton molecules above $T_c$
by continuing the Nambu-Jona-Lasinio description from below
$T_c$ upwards.

Our present discussion
of mesonic bound states  should not be confused with
quasi-hadronic states  found in early lattice
calculations\cite{detar85} for quarks and antiquarks propagating
in the space-like direction. Their spectrum, known as ``screening
masses'' is generated mostly by ``dynamical confinement" of the
spatial Wilson loop which is a nonperturbative phenomenon seen via
the lattice calculations. Similar effects will be given here by
the instanton molecule interaction.

In this work we are able to construct a smooth transition from the chirally
broken to the chirally restored sector in terms of continuity in
the masses of the $\sigma$ and $\pi$ mesons, vanishing at
$T\rightarrow T_c$. 
The crucial part
of strong binding in
our picture of $\bar q q$ mesons (or molecules) is
the quasi-local interaction due to instanton molecules
(the ``hard glue'').
We found that the tight binding of these mesons near $T_c$ enhences
the wave function at the origin, and gives us
additional understanding of the nonperturbative
hard glue (epoxy) which is preserved
at $T>T_c$.

In our Nambu-Jona Lasinio formalism we construct giant collective vibrations 
with an essentially SU(4) structure. There giant vibrations were found
in lattice calculations by by Asakawa et al.\cite{asakawa03} at
two temperatures.
The lower temperature was in the region of the Shuryak-Zahed
$T_{zb}$, and confirmed that the binding energy was very small. 
The higher temperature was at $\sim 500$ MeV, above those reached
by RHIC, and the lattice calculations suggested perturbative quasiparticles,
indicating production of the QGP at this temperature.

\ack
GEB and ES were partially supported by the
US Department of Energy under Grant No. DE-FG02-88ER40388.
CHL is supported by Korea Research Foundation
Grant (KRF-2002-070-C00027).

\end{document}